\documentclass[10pt,aps,prd,twocolumn,nofootinbib,showpacs,showkeys,superscriptaddress]{revtex4-2}

\usepackage{multirow} 
\usepackage{makecell} 
\usepackage{psfrag} 
\usepackage{mathrsfs} 
\usepackage{amssymb, bm} 
\usepackage{amsmath,amsthm} 
\usepackage{epstopdf} 
\usepackage[breaklinks=true]{hyperref} 
\usepackage{enumerate} 
\usepackage{longtable} 
\usepackage{subfigure} 
\usepackage{color} 
\usepackage{mathrsfs} 
\usepackage{graphicx} 
\usepackage{bm,natbib,url,textcase} 
\usepackage{xurl} 
\usepackage[title]{appendix}

\usepackage[english]{babel} \usepackage[T1]{fontenc} \usepackage{comment} 
\numberwithin{equation}{section}

\usepackage{geometry}
\allowdisplaybreaks

\hypersetup{
}

\usepackage{tikz}

\newcommand{\be}{\begin{equation}} 
\newcommand{\ee}{\end{equation}}

\definecolor{purple}{rgb}{1,0,1} \definecolor{lime}{HTML}{a6CE39} 

\newcommand{\orcidicon}{%
	\begin{tikzpicture}

		\draw[lime, fill=lime] (0,0) circle [radius=0.15] 
		node[white] {{\fontfamily{qag}\selectfont \tiny ID}}; 
		\draw[white, fill=white] (-0.0625,0.095) circle 
		[radius=0.007];
	\end{tikzpicture} \hspace{-2mm} }

\newcommand\orcidValerio{{\href{https://orcid.org/0000-0002-2601-1870}{\orcidicon}}}

\DeclareMathOperator{\arctanh}{arctanh}

\begin{document} \def\theequation{\arabic{section}.\arabic{equation}}

\title{The thermal view of singularity-free scalar-tensor spacetimes}



\author{Valerio Faraoni\orcidValerio} \email[]{vfaraoni@ubishops.ca} 
\affiliation{Department of Physics \& Astronomy, Bishop's University, 2600 
College Street, Sherbrooke, Qu\'ebec, Canada J1M~1Z7}

\author{Nikki Veilleux} \email[]{nveilleux21@ubishops.ca} 
\affiliation{Department of Physics \& Astronomy, Bishop's University, 2600 
College Street, Sherbrooke, Qu\'ebec, Canada J1M~1Z7}

\begin{abstract}

The two-parameter inhomogeneous and time-dependent Pimentel solution of 
Brans-Dicke theory is analyzed to probe and extend the new thermal 
view in which 
General Relativity is the zero-temperature (equilibrium) state of 
scalar-tensor gravity. As the parameters vary, we uncover phenomenology 
not found before with other exact solutions, nor contemplated thus far in 
the general theory. In the process, we also discuss the anomalous limit to 
General Relativity of the Pimentel geometry and show how the Mars 
solution of the Einstein equations is its Einstein frame version, 
elucidating the relation between these geometries.

\end{abstract}


\maketitle

\section{Introduction} \label{sec:1} \setcounter{equation}{0}

Although General Relativity (GR) has been very successful, there is little 
doubt that it cannot be the ultimate theory of gravity. GR predicts its 
own failure at spacetime singularities, such as the Big Bang in cosmology 
and the singularities hidden inside black hole horizons. It is expected 
that quantum mechanics will eventually cure this problem but, as soon as 
first order quantum corrections are introduced, they ``break'' GR by 
introducing extra degrees of freedom, raising the order of the field 
equations, and introducing extra terms in the Einstein-Hilbert action 
\cite{Stelle:1976gc,Stelle:1977ry}.  In particular, early universe 
inflation was first realized by Starobinsky by introducing quadratic 
curvature terms to quantum-correct GR \cite{Starobinsky}.  Corrections 
quadratic in the Ricci scalar $R$ can be described with an equivalent 
massive scalar field. More generally, consider metric $f(R)$ gravity 
described by the action\footnote{We follow the notation of 
Ref.~\cite{Wald:1984rg}, using units in which the speed of light $c$ and 
Newton's constant $G$c are unity, except in Secs.~\ref{sec:2} 
and~\ref{sec:3} where, for ease of comparison with 
Refs.~\cite{Mars:1995jv,Pimentel:1996du}, we use $8\pi G=1$.} 
\be 
S_{f(R)}= \int d^4 x \, \sqrt{-g} \left[ \frac{f(R)}{16 \pi} +{\cal 
L}^\mathrm{(m)} \right] \,, 
\ee 
which contains as a special case the 
Starobinsky Lagrangian $f(R)=R+\alpha R^2$. Here $g$ is the determinant of 
the metric tensor $g_{\mu\nu}$, $R\equiv g^{\mu\nu} R_{\mu\nu}$ is the 
Ricci scalar, $R_{\mu\nu} $ is the Ricci tensor of $g_{\mu\nu}$, $f(R)$ is 
a non-linear function of the Ricci scalar, and ${\cal L}^\mathrm{(m)}$ is 
the matter Lagrangian density.  It is well known (see 
Refs.~\cite{Sotiriou:2008rp,DeFelice:2010aj,Nojiri:2010wj} for reviews) 
that metric $f(R)$ gravity is equivalent to a Brans-Dicke theory with 
Brans-Dicke coupling parameter $\omega=0$, scalar field $\phi=f'(R)$, and 
scalar field potential \be V(\phi) = R f'(R)- f(R) \Bigg|_{\phi=f'(R)} \,. 
\ee

Similarly, the low-energy limit of the simplest string theory (bosonic 
string theory) yields $\omega=-1$ Brans-Dicke gravity instead of GR 
\cite{Callan:1985ia,Fradkin:1985ys}.

Independent motivation for the study of theories of gravity alternative to 
GR comes from the 1998 discovery of the acceleration of the cosmic 
expansion, originally attributed to a cosmological constant $\Lambda$ or 
to a mysterious dark energy \cite{AmendolaTsujikawabook}.  If confirmed, 
the recent DESI results \cite{DESI:2024aax,DESI:2024uvr,DESI:2024lzq} 
would exclude the cosmological constant as an explanation. Dynamical dark 
energy is still a possibility, but it is completely {\em ad hoc}: it was 
introduced almost overnight only to explain the new data, and it is akin 
to a fudge factor. For this reason, many cosmologists have resorted to 
modified gravity as an alternative to dark energy 
\cite{Capozziello:2003tk,Carroll:2003wy}. In this approach, one does not 
invoke dark energy and gravity deviates from GR at the largest 
(cosmological) scales. There is proof of principle that alternative 
gravity can explain the cosmic acceleration, and the class of $f(R)$ 
theories is particularly popular for this purpose 
\cite{Sotiriou:2008rp,DeFelice:2010aj,Nojiri:2010wj}.

$f(R)$ gravity is a subclass of the more general scalar-tensor gravity, 
which began with Brans-Dicke theory \cite{Brans:1961sx}, was later 
generalized to ``first-generation'' scalar-tensor gravity 
\cite{Bergmann:1968ve,Nordtvedt:1968qs, Wagoner:1970vr,Nordtvedt:1970uv}, 
and finally extended to Horndeski \cite{Horndeski} and DHOST gravity (see 
\cite{Kobayashi:2019hrl,DHOSTreview1,DHOSTreview2} for reviews).

The (Jordan frame) Brans-Dicke action is 
\begin{eqnarray} 
S_\mathrm{BD} 
&=& \int d^4 x \, \frac{ \sqrt{-g}}{16\pi}  \left[ \phi R 
-\frac{\omega}{\phi} \, 
g^{\mu\nu} \, \nabla_{\mu}\phi \nabla_{\nu}\phi -V(\phi) \right. 
\nonumber\\
&&\nonumber\\
&\, & \left. + {\cal L}^\mathrm{(m)} \right] \,, \label{BDaction}
\end{eqnarray} 
where $\phi$ is the Brans-Dicke scalar field, $V(\phi)$ is its 
potential, and the constant $\omega$ is the ``Brans-Dicke coupling''. 
The field equations obtained by varying the action~(\ref{BDaction}) with 
respect to the inverse metric $g^{\mu\nu}$ and to $\phi$ are 
\begin{eqnarray}
    G_{\mu\nu} &=& \frac{8\pi}{\phi} T_{\mu\nu}^\textrm{(m)}
+  T_{\mu\nu}^{(\phi)} \nonumber\\ &&\nonumber\\ &=& \frac{8\pi}{\phi} 
  T_{\mu\nu}^\textrm{(m)}
 + \frac{\omega}{\phi^2} \left( \nabla_\mu \phi \nabla_\nu \phi 
   -\frac{1}{2} \, g_{\mu\nu} \, \nabla_\lambda\phi \nabla^\lambda\phi 
   \right) \nonumber\\
&&\nonumber\\
&\, & +
\frac{1}{\phi} \left( \nabla_\mu \nabla_\nu \phi - g_{\mu\nu} \Box \phi 
\right) - \frac{V(\phi)}{2\phi} \, g_{\mu\nu} \,,\label{BDfe1} 
\end{eqnarray}

\begin{equation}
    \Box \phi = \frac{1}{2\omega +3} \left[ 8\pi T^\textrm{(m)}
 + 2V(\phi) - \phi \, V' (\phi) \right]\,,\label{BDfe2} 
\end{equation} 
where $G_{\mu\nu} \equiv R_{\mu\nu}- g_{\mu\nu} \, R/2$ is the Einstein  
tensor, $T_{\mu\nu}^\textrm{(m)} $ is the matter stress-energy tensor,  
$T^\textrm{(m)} \equiv g^{\mu\nu} T_{\mu\nu}^\textrm{(m)} $, and $\Box  
\equiv g^{\alpha\beta} \nabla_{\alpha} \nabla_{\beta}$.  We assume  
$2\omega+3>0$ to ensure that the Brans-Dicke field $\phi$ is not a  
phantom \cite{Faraoni:2004pi}.

Brans-Dicke gravity can be reformulated in the Einstein conformal frame:  
the conformal transformation and scalar field redefinition \be 
g_{\mu\nu}\to \tilde{g}_{\mu\nu}=\phi \, g_{\mu\nu} \,, 
\ee 
\be 
d\tilde{\phi}= \sqrt{\frac{2\omega+3}{16\pi}} \, \frac{d\phi}{\phi} \,, 
\ee 
bring the action~({\ref{BDaction}) to the form 
\begin{eqnarray} 
S_\mathrm{BD} &=& \int d^4 x \, \sqrt{-\tilde{g}} \left[ \frac{ 
\tilde{R}}{16\pi}  
-\frac{1}{2} \, \tilde{g}^{\mu\nu} \, \tilde{\nabla}_{\mu} \tilde{\phi} 
\tilde{\nabla}_{\nu} \tilde{\phi} -U( \tilde{\phi})\right.\nonumber\\
&&\nonumber\\
&\, & \left.  + \, \mbox{e}^{ - 8\sqrt{ \frac{\pi}{2\omega+3} } \, 
\tilde{\phi} }\, {\cal L}^\mathrm{(m)} \right]  \,, 
\end{eqnarray} 
where the theory looks like GR but an ever-present scalar field couples 
explicitly to matter.

Recently, a new picture of scalar-tensor gravity based on an analogy with 
heat dissipation in an imperfect fluid was introduced. There are two basic 
ideas: 1)~the field equations of scalar-tensor gravity are written as 
effective Einstein equations, with the extra (non-matter) terms grouped in 
an effective stress-energy tensor. The latter has the form of a 
dissipative fluid energy-momentum tensor (this structure is common to all 
symmetric two-index tensors and has no physical meaning in itself 
\cite{Faraoni:2023hwu}).
2)~This {\it effective} fluid satisfies the basic postulate of Eckart's 
 thermodynamics of dissipative fluids \cite{Eckart:1940te}, which is a 
 constitutive relation generalizing the Fourier law of heat conduction. 
 This is a little miracle that makes it possible to introduce a concept of 
 ``temperature of gravity'' relative to GR. In this picture, GR is the 
 zero-temperature state of thermal equilibrium, while scalar-tensor 
 gravity is an excited state at positive temperature.

Now to the specifics: assuming that the scalar field gradient 
$\nabla^{\mu}\phi$ is timelike and future-oriented (otherwise it is not 
possible to apply the formalism), the four-velocity of the effective 
$\phi$-fluid is defined by \be u^{\mu} \equiv \frac{\nabla^{\mu} \phi}{ 
\sqrt{-\nabla_{\nu} \phi \nabla^{\nu} \phi}} \,.\label{4-velocity} \ee The 
effective stress-energy tensor $T_{\mu\nu}^{(\phi)}$ in Eq.~(\ref{BDfe1}) 
has the dissipative form \cite{Eckart:1940te} 
\be 
T_{\mu\nu} =\rho u_{\mu} 
u_{\nu} +P h_{\mu\nu} +\pi_{\mu\nu} + q_{\mu} u_{\nu} + q_{\nu} u_{\mu} 
\,, \label{eq:imperfect} 
\ee 
where observers comoving with the 
$\phi$-fluid and defined by their four-velocity $u^{\mu}$ experience $ 
h_{\mu\nu} \equiv g_{\mu\nu} + u_{\mu} u_{\nu} $ as the Riemannian metric 
of 3-space.  $\rho, P, q_{\mu}$, and $\pi_{\mu\nu}$ are, respectively, the 
effective energy density, effective pressure, effective heat flux density, 
and effective anisotropic stress tensor of the $\phi$-fluid 
\cite{Ellis:1971pg}. They are computed in \cite{Faraoni:2018qdr} for 
first-generation scalar-tensor gravity and in 
\cite{Quiros:2019gai,Giusti:2021sku} for ``viable'' Horndeski gravity.

Eckart's first order thermodynamics of dissipative fluids 
\cite{Eckart:1940te} relies on the generalized Fourier law 
\be 
q_{\mu} 
=-{\cal K} h_{\mu\nu} \left( \nabla^{\nu} {\cal T} +{\cal T} \dot{u}^{\nu} 
\right) \,,\label{FourierEckart} 
\ee 
where ${\cal T}$ and ${\cal K}$ are 
the fluid's temperature and thermal conductivity and $ \dot{u}^{\mu} 
\equiv u^{\beta} \nabla_{\beta} u^{\mu}$ is its four-acceleration. 
Although subject to fundamental limitations 
\cite{Maartens:1996vi,Andersson:2006nr}, Eckart's theory is widely used as 
a simple model for dissipative fluids.

{\em A priori}, the effective $\phi$-fluid should not know about the 
Eckart-Fourier law~(\ref{FourierEckart}), contrary to real fluids. 
However, the direct computation of $q_{\mu}^{(\phi)}$ and $\dot{u}^{\mu}$ 
\cite{Faraoni:2018qdr,Faraoni:2021lfc,Faraoni:2021jri,Giusti:2021sku} 
gives the surprising result \be
 q_{\mu}^{(\phi)} =-{\cal K}{\cal T} \, \dot{u}_{\mu} \,, \ee from which 
one reads off \be {\cal K}{\cal T} = \frac{ \sqrt{-\nabla^{\mu} \phi 
\nabla_{\mu} \phi} }{ 8\pi \phi} \label{KTdefinition} \ee as the 
coefficient of the acceleration, which defines a ``temperature of 
gravity'' ${\cal T}$ and an effective ``thermal conductivity of 
spacetime'' ${\cal K}$ (unfortunately, only their product is known). It is 
a temperature relative to GR, which is the trivial case $\phi =$~const. of 
scalar-tensor gravity. Setting $\phi=$~const. produces
 ${\cal K}{\cal T} = 0$. Naturally, the excitation of the propagating 
scalar degree of freedom promotes gravity to a ``thermally excited'' state 
with ${\cal K}{\cal T}>0$.

The approach of scalar-tensor gravity to GR is described by decreasing 
${\cal K}{\cal T} $ (and, eventually, GR is obtained if ${\cal K}{\cal 
T}\to0$) and its departure from GR by increasing ${\cal K}{\cal T}$. 
Whether gravity approaches GR or departs from it is described by the 
evolution of ${\cal K}{\cal T}$ in the comoving time $\tau$ of the 
effective fluid. This evolution obeys the law 
\cite{Faraoni:2018qdr,Faraoni:2021lfc,Faraoni:2021jri,Giusti:2021sku} 
\be 
\frac{d\left( {\cal K}{\cal T}\right)}{d\tau} = 8\pi \left( {\cal K}{\cal 
T}\right)^2 -\Theta {\cal K}{\cal T} +\frac{ \Box\phi}{8\pi \phi} \,, 
\label{evolution_general} 
\ee 
where $\Theta \equiv \nabla_{\mu} u^{\mu}$ 
is the expansion scalar of the $\phi$-fluid. In first-generation 
scalar-tensor gravity, Eq.~(\ref{evolution_general}) becomes 
\begin{eqnarray} 
\frac{d \left( {\cal K}{\cal T}\right)}{d\tau} &=& 8\pi 
\left( {\cal K}{\cal T}\right)^2 -\Theta {\cal K}{\cal T}
+ \frac{ T^\mathrm{(m)} }{\left( 2\omega + 3 \right) \phi} \nonumber\\ 
  &&\nonumber\\
&\, & +\frac{1}{8\pi
\left( 2\omega + 3 \right)} \left( V' -\frac{2V}{\phi} 
-\frac{d\omega}{\phi d\phi} \, \nabla^{\alpha}\phi \nabla_{\alpha} \phi 
\right) \,. \nonumber\\
&& \label{evolution_general2}
\end{eqnarray} Positive terms in the right-hand side ``heat up'' gravity 
(i.e., move it away from GR), while negative terms ``cool'' it (i.e., move 
it toward GR).

For vacuum Brans-Dicke gravity \cite{Brans:1961sx} with $\omega=$~const. 
and no potential (the case relevant for this work), 
Eq.~(\ref{evolution_general2})  reduces to 
\be 
\frac{ d \left({\cal 
K}{\cal T} \right)}{d\tau} = {\cal K}{\cal T} \left( 8\pi {\cal K}{\cal 
T}-\Theta \right) \,.
  \label{evolution-reduced} 
\ee 
The $ \left( \Theta, {\cal K}{\cal T} 
\right)$ plane turns out to be very useful to represent the ``thermal'' 
evolution of scalar-tensor spacetimes when $\Box\phi=0$ and the simplified 
evolution equation~(\ref{evolution-reduced}) holds \cite{Faraoni:2025alq}.

The basic picture is the following: if the expansion $\Theta $ is 
negative, $d\left( {\cal K}{\cal T}\right)/d\tau$ is positive and gravity 
``heats up''. The critical half-line $8\pi {\cal K}{\cal T}=\Theta$ 
divides the remaining region $\Theta>0$ in two: if $8\pi {\cal K}{\cal 
T}>\Theta$, gravity ``heats up'' and moves away from GR. If, 
instead, $8\pi {\cal K}{\cal T}<\Theta$, gravity ``cools'' moving toward 
GR. Fixed points of the evolution $\left( \Theta, 
{\cal K}{\cal T} \right) = \left( \Theta_0 , \frac{\Theta_0 }{8\pi} 
\right)$ with $\Theta_0=$~const. lie along the critical half-line. This 
line can be crossed dynamically at other points. These distinct thermal 
behaviours are explained 
physically by the dominance of the scalar degree of freedom $\phi$ over 
the two massless spin two degrees of freedom of GR, or {\it vice-versa} 
\cite{Faraoni:2025alq}, and shed light on the old problem of the 
convergence of scalar-tensor gravity to GR in the early universe 
\cite{Damour:1992kf,Damour:1993id}, or its departure from it 
\cite{Serna:2002fj}.

The thermal description of scalar-tensor gravity has proved quite useful 
in unifying results scattered in the literature and apparently 
disconnected, providing a rather comprehensive framework still under 
development 
\cite{Faraoni:2018qdr,Faraoni:2021lfc,Faraoni:2021jri,Giusti:2021sku, 
Giusti:2022tgq,Giardino:2022sdv}. It contains several 
ideas (tested on exact solutions of 
several scalar-tensor theories \cite{Faraoni:2022doe, 
Faraoni:2022jyd,Giardino:2023qlu, Faraoni:2022fxo,Houle:2024sxs, 
Karolinski:2024nwp}), also in the Einstein conformal frame where 
temperature is traded with chemical potential \cite{Faraoni:2022gry}. The 
present work continues this analysis using the Pimentel solution of 
Brans-Dicke gravity, which reveals itself quite useful and provides 
insight on new thermal behaviours not discovered before. We begin by 
discussing a stiff fluid solution of GR that eventually turns out to be 
the Einstein frame version of the Pimentel geometry.

\section{Mars solution of the Einstein equations} 
\label{sec:2} 
\setcounter{equation}{0}

The Mars solution of the Einstein equations \cite{Mars:1995jv} is 
cylindrically symmetric, time-dependent, inhomogeneous, globally 
hyperbolic, geodesically complete and singularity-free, and it contains a 
Bianchi~II anisotropic universe as a special case \cite{Mars:1995jv}.  
The original interest arose with the realization that inhomogeneity can  
 prevent singularities without violations of the energy conditions 
\cite{Senovilla:1990rt,Chinea:1992nq,Ruiz:1992np}.

The Mars solution is sourced by a stiff fluid. Its line element is 
\begin{eqnarray} d\tilde{s}^2 &=& \, \mbox{e}^{\alpha_0 r^2} \cosh\left( 
2pt 
\right) \left( -dt^2+dr^2 \right) +r^2 \cosh \left(2pt\right) d\varphi^2 
\nonumber\\
&&\nonumber\\
&\, & +\frac{ \left( dz
+pr^2 d\varphi \right)^2}{\cosh\left( 2pt \right)} \label{1} 
\end{eqnarray} in cylindrical coordinates, where $ -\infty< t <+\infty$, $ 
r\geq 0 $, $0\leq \varphi < 2\pi $, $ -\infty<z< +\infty$, and where 
$\alpha_0$ and $p\neq 0$ are constants.\footnote{Mars \cite{Mars:1995jv} 
uses the symbol $a$ instead of $p$. Here we follow the notation of 
\cite{Pimentel:1996du}, to which we refer extensively. The reason for 
using tildes in this section will be clear in Sec.~\ref{sec:3}, where it 
is shown that the Mars geometry is the Einstein frame version of the 
Pimentel solution of Brans-Dicke gravity.} $p^{-1}$ is a length or time 
scale, $[ p ] = [ L^{-1} ]$, while the dimensions of $s$ are $[ s ] = 
[L^{-2} ]$.

The $z$-axis is a symmetry axis and the metric components 
$g_{\mu\nu}(t,r)$ do not depend on $\varphi$ or $z$. The solution is also 
symmetric under time reversal $t\to -t$.

The Mars solution is sourced by a perfect fluid with stiff equation of 
state \be \tilde{P}=\tilde{\rho}= \frac{ p^2 \left( \alpha_0 -1 \right) \, 
\mbox{e}^{-\alpha_0 p^2 r^2} }{\cosh\left( 2pt \right)} \,, \ee where 
$\tilde{\rho}$ and $\tilde{P}$ are the fluid's energy density and 
pressure, respectively.  We choose $\alpha_0>1$ to ensure a positive 
energy density.  We adopt Pimentel's notation $s\equiv \alpha_0 p^2$, then 
the condition $\alpha_0>1$ and the line element read \be 0 \leq p^2 <s 
\,,\label{4} \ee \begin{eqnarray} d\tilde{s}^2 &=& \, \mbox{e}^{sr^2} 
\cosh\left( 2pt \right) \left( -dt^2+dr^2 \right) +r^2 \cosh 
\left(2pt\right) d\varphi^2 \nonumber\\
&&\nonumber\\
&\, & +\frac{ \left( dz +pr^2 d\varphi \right)^2}{\cosh\left( 2pt
\right)} \,,\label{3} \end{eqnarray} while \be \tilde{P}=\tilde{\rho}= 
\frac{ \left( s -p^2 \right) \, \mbox{e}^{-s r^2} }{\cosh\left( 2pt 
\right)} \,. \label{5} \ee

The four-velocity of the fluid $\tilde{u}^{\mu} \equiv dx^{\mu}/d\tau$ 
(where $\tau$ is the fluid's proper time) has components 
\begin{equation}
    \tilde{u}^\mu = \left( \frac{\mbox{e}^{-sr^2/2} }{ \sqrt{\cosh(2pt)}}, 
0, 0, 0 \right) \label{6} 
\end{equation} 
and is normalized, 
$\tilde{g}_{\mu\nu} \tilde{u}^{\mu} \tilde{u}^{\nu}=-1 $.  Its 
four-acceleration is 
\begin{equation}
    \tilde{a}^\mu \equiv \tilde{u}^\alpha\tilde{\nabla}_\alpha 
\tilde{u}^\mu = \left( 0, \frac{sr \, \mbox{e}^{-sr^2/2}}{\cosh(2pt)}, 0, 
0 \right) \,,\label{8} 
\end{equation} 
while the expansion scalar is 
\begin{equation}
    \Theta \equiv \tilde{\nabla}_\mu \tilde{u}^\mu = p \; \frac{ 
\mbox{e}^{-sr^2/2} \, \sinh(2pt)}{\cosh^{3/2}(2pt)} \,.\label{9} 
\end{equation} 
The expansion vanishes asymptotically in the infinite past 
and future, $\tilde{\Theta} \to 0$ as $t\to \pm \infty$.  The fluid 
contracts (expansion scalar $\tilde{\Theta}<0$) for any $p\neq 0$ for all 
$t<0$, stops ($\tilde{\Theta}=0$) at $t=0$, and then re-expands 
($\tilde{\Theta}>0$) for all $t>0$, without Big Bang-like singularities.  
This behaviour resembles a cosmological bounce, which Mars attributes to 
the inhomogeneity since the weak and null energy conditions are not 
violated (assuming $\alpha_0>1$). The discovery of the Mars solution 
followed that of similar geodesically complete solutions 
\cite{Senovilla:1990rt,Chinea:1992nq,Ruiz:1992np}.

The Mars geometry describes a 2-parameter $\left(\alpha_0, p \right) $ 
family of solutions of the Einstein equations, which reduces to the 
Minkowski geometry for $ \left(\alpha_0, p \right)=\left(0, 0 \right) $. 
Regardless of $\alpha_0$, in the limit $p \to 0$ the solution becomes 
static (with $\tilde{\Theta}=0$) and diagonal, 
\begin{eqnarray} 
d\tilde{s}^2_{(0)} &=& \, \mbox{e}^{sr^2} \left( -dt^2+dr^2 \right) +r^2
 d\varphi^2 + dz^2 \,, 
\end{eqnarray} 
while 
\be 
\tilde{P}=\tilde{\rho}=s\, 
\mbox{e}^{-sr^2} \,. 
\ee 
This limit is obtained keeping $\alpha_0 \, p^2=s$ constant.

The shear tensor of $\tilde{u}^{\mu}$ has the only non-vanishing 
components \be \tilde{\sigma}_{11}= \tilde{\sigma}_{22}= - 
\frac{\tilde{\sigma}_{33} }{2} = \frac{2\tilde{\Theta}}{3} \ee and the 
vorticity tensor vanishes identically, $\tilde{\omega}_{\mu\nu}=0$ 
\cite{Mars:1995jv}.

\subsection{Mars solution as a scalar field solution of GR}

Here we show that the Mars geometry~(\ref{3}) corresponds to a scalar 
field solution of the Einstein equations and we find explicitly the 
corresponding scalar field. This reformulation is crucial to relate the 
Mars geometry with the Pimentel solution of Brans-Dicke gravity in the 
next section.
 
It is well known that a stiff fluid corresponds to a free scalar field, 
provided that its gradient is timelike and future-oriented. Therefore, the 
Mars solution sourced by a stiff fluid can be seen as a scalar field 
solution. The stress-energy tensor of a minimally coupled scalar field 
$\tilde{\phi}$ is \be \tilde{T}_{\mu\nu}^{ (\tilde{\phi})} 
=\tilde{\nabla}_{\mu} \tilde{\phi} \, \tilde{\nabla}_{\nu}\tilde{\phi} 
-\frac{1}{2} \, \tilde{g}_{\mu\nu} \, \tilde{\nabla}^{\alpha}\tilde{\phi} 
\tilde{\nabla}_{\alpha}\tilde{\phi}
+ V ( \tilde{\phi} ) \, \tilde{g}_{\mu\nu} \,,\label{12} \ee where $V ( 
  \tilde{\phi} )$ is the scalar field potential. Assuming 
  $\tilde{\nabla}^{\mu} \tilde{\phi}$ to be timelike and future-oriented, 
  the scalar field is equivalent to an effective fluid with four-velocity 
  \begin{equation}
    \tilde{u^\mu} = \frac{\tilde{\nabla}^{\mu} 
\tilde{\phi}}{\sqrt{-\tilde{g^{\alpha \beta}}\tilde{\nabla}_\alpha 
\tilde{\phi} \, \tilde{\nabla}_\beta \tilde{\phi}}} \label{13} 
\end{equation} and then the scalar field stress-energy tensor~(\ref{12}) 
assumes the perfect fluid structure \be \tilde{T}_{\mu\nu}^{ 
(\tilde{\phi})} = \left( \tilde{\rho}+\tilde{P} \right) \tilde{u}_{\mu} 
\tilde{u}_{\nu} +\tilde{P} \, \tilde{g}_{\mu\nu} = \tilde{\rho} \, 
\tilde{u}_{\mu} \tilde{u}_{\nu} +\tilde{P} \, \tilde{h}_{\mu\nu} \,, 
\label{12} \ee where \be \tilde{h}_{\mu\nu} = \tilde{g}_{\mu\nu} + 
\tilde{u}_{\mu} \tilde{u}_{\nu} \label{15} \ee is the Riemannian metric 
seen by the observers with four-velocity $\tilde{u}^{\mu}$. The 
$\phi$-fluid has effective energy density \begin{equation} \tilde{\rho} = 
\tilde{T}_{\mu \nu}^{(\tilde{\phi})} \, \tilde{u}^\mu \tilde{u}^\nu = 
-\frac{1}{2} \, \tilde{g}^{\alpha \beta} \tilde{\nabla}_{\alpha} 
\tilde{\phi} \, \tilde{\nabla}_{\beta}\tilde{\phi}
+ V(\tilde{\phi})  \label{16} \end{equation} and effective pressure 
  \begin{equation} \tilde{P} = \frac{1}{3} \, \tilde{h}^{\mu\nu} \, 
  \tilde{T}_{\mu
\nu}^{(\tilde{\phi})} =
 {-\frac{1}{2} \, \tilde{g}^{\alpha \beta} \tilde{\nabla}_\alpha 
\tilde{\phi} \, \tilde{\nabla}_\beta\tilde{\phi} - V(\tilde{\phi})} \,. 
\label{17}
 \end{equation} For a free scalar field, $V ( \tilde{\phi} ) =0$ yields 
$\tilde{P}=\tilde{\rho}$ and this field effectively behaves as a stiff 
fluid. The continuity equation $\tilde{\nabla}^{\nu} 
\tilde{T}_{\mu\nu}^{(\tilde{\phi} )} =0$ becomes the Klein-Gordon equation 
\be \tilde{\Box} \tilde{\phi} = \frac{1}{ \sqrt{ - \tilde{g}} } \, 
\partial_{\mu} \left( \sqrt{ - \tilde{g}} \, \tilde{g}^{\mu\nu} 
\partial_{\nu} \tilde{\phi} \right) =0 \,. \ee


Combining Eq.~(\ref{5}) with Eqs.~(\ref{16}) or~(\ref{17}) with $V=0$ 
gives (an overdot denoting differentiation with respect to the coordinate 
time) \be \frac{ (s-p^2) \, \mbox{e}^{-sr^2} }{\cosh(2pt)} = - \frac{1}{2} 
\, \tilde{g}^{00} \left( \dot{ \tilde{\phi}} \right) ^2 \label{22} \ee if 
we assume $\tilde{\phi}=\tilde{\phi}(t)$, then \be \left( 
\dot{\tilde{\phi}} \right)^2 =2(s-p^2) \,, \ee which has the solution 
\begin{equation}
    \tilde{\phi}(t) = \tilde{\phi}_0 \, t + \tilde{\phi}_1 = \pm 
\sqrt{2(s-p^2)} \,t + \tilde{\phi}_1 \, , \end{equation} where 
$\tilde{\phi}_1$ is an integration constant and $\tilde{\phi}_0^2 \equiv 
2(s-p^2)$.  We can set $\tilde{\phi}_1=0$ since it does not contribute to 
$\tilde{\rho}=\tilde{P}$.

In principle $\tilde{\phi}_0 = \pm \sqrt{ 2(s-p^2)} $ but, in order to 
keep $\tilde{\nabla}^{\mu}\tilde{\phi}$ future-oriented, we take the 
negative root. Then, 
\begin{equation}
    \tilde{\nabla}^\mu \tilde{\phi} = \tilde{g}^{\mu \nu} 
\tilde{\nabla}_\nu \tilde{\phi} = {\delta^{\mu} }_0 \, \tilde{\phi}_0 \, 
\tilde{g}^{00} = - \frac{ {\delta^{\mu} }_0 \, \tilde{\phi}_0 \, 
\mbox{e}^{-sr^2} }{\cosh(2pt)} 
\end{equation} 
and $\tilde{\nabla}^0 
\tilde{\phi} >0$ if $\tilde{\phi}_0 < 0 $, therefore 
\be 
\tilde{\phi}_0 = 
-\sqrt{2(s-p^2)} \,.\label{24} 
\ee 
Although the Mars line element is 
time-symmetric, the scalar field is not. However, only $ \left( 
d\tilde{\phi}/dt\right)^2$ enters the expression of 
$\tilde{P}=\tilde{\rho}$, which is invariant under time reversal $t\to 
-t$.

Now the vorticity $\tilde{\omega}_{\mu\nu}$ is seen to vanish identically 
(in agreement with \cite{Mars:1995jv}) because $\tilde{u}^{\mu}$ is 
derived from a scalar field gradient, while the normalized 
four-velocity~(\ref{13})  of the effective fluid \begin{equation}
    \tilde{u}^\mu = \left( \frac{ \mbox{e}^{-sr^2/2}}{\sqrt{\cosh(2pt)}}, 
0, 0, 0 \right) \end{equation} matches Mars' four-velocity~(\ref{6}).

\section{Pimentel solution of Brans-Dicke gravity} 
\label{sec:3} 
\setcounter{equation}{0}

Pimentel \cite{Pimentel:1996du} proposed the solution of vacuum 
Brans-Dicke gravity (with $\omega=$~const., $V(\phi) \equiv 0$)  
\begin{eqnarray} d\tilde{s}^2 &=& \mbox{e}^{ht} \left[ \mbox{e}^{sr^2} 
\cosh\left( 2pt \right) \left( -dt^2+dr^2 \right) +r^2 \cosh 
\left(2pt\right) d\varphi^2 \right. \nonumber\\
&&\nonumber\\
&\, & \left.
+\frac{ \left( dz +pr^2 d\varphi \right)^2}{\cosh\left( 2pt \right)} 
\right] \,,\label{31} 
\end{eqnarray} 
\be 
\phi(t) = \phi_0 \, 
\mbox{e}^{-ht} \,,\label{32} 
\ee 
where $h, s, p$, and $\phi_0>0$ are 
constants satisfying\footnote{Note a typographical error in 
Ref.~\cite{Pimentel:1996du}: there, $(s-p)$ appears in the bracket on the 
right-hand side instead of $(s-p^2)$, which is dimensionally incorrect 
because $[s]=[L^{-2}]$ and $[p]=[L^{-1}]$.} \be h^2 \left( 
2\omega+3\right) =4(s-p^2) \,, \ee and where $h^2 > 0$ ensures that $ 0 < 
p^2 < s$. The coordinates vary in the range $ -\infty<t<+\infty$, $r\geq 
0$, $0\leq \varphi <2\pi$, $-\infty <z<+\infty$. This solution is 
cylindrically symmetric, time-dependent, and has no spacetime 
singularities. $\omega $ is a parameter of the theory, while $s$ and $p$ 
are parameters of the solution. The Pimentel solutions form a 2-parameter 
family conformal to the Mars solution of GR, therefore the causal 
structure is the same. Pimentel derives his solution by solving directly 
the (Jordan frame) Brans-Dicke field equations. However, the Mars solution 
is nothing but the Einstein frame version of Pimentel's solution. In fact, 
take the usual map from Jordan to Einstein frame of Brans-Dicke gravity 
\be 
g_{\mu\nu} \to \tilde{g}_{\mu\nu} =\phi g_{\mu\nu} \,, \quad\quad 
\tilde{\phi}= \sqrt{\frac{2\omega+3}{2}} \,\ln \left( \frac{ \phi}{ 
\mbox{const.}} \right)  \label{36} 
\ee 
(in units in which $8\pi G=1$). 
Regarding the scalar field $\tilde{\phi}(t) = -\sqrt{4(s-p^2)} \, t $ that 
we derived for the Mars solution as the Einstein frame scalar field, we 
obtain 
\be 
\phi(t)=\phi_0 \exp \left( -\sqrt{ \frac{4(s-p^2) }{2\omega+3}} 
\, t \right) \equiv \phi_0 \, \mbox{e}^{-ht} \,, \label{38} 
\ee 
where 
\begin{equation}
    h^2 = \frac{ 4(s-p^2)}{2\omega+3} >0 \,,\label{33} 
\end{equation} 
which reproduces the scalar field of the Mars solution derived in 
Sec.~\ref{sec:2} (remember that Mars considers only a stiff fluid and does 
not discuss scalar fields).

The Pimentel line element obtained from the conformal mapping of the (now 
Einstein frame) Mars solution back to the Jordan frame is 
\begin{eqnarray} 
ds^2 &=& \frac{ d\tilde{s}^2}{\phi} = \frac{ \mbox{e}^{ht}}{\phi_0} \left[ 
\mbox{e}^{sr^2} \cosh(2pt) \left(-dt^2+dr^2\right) \right.\nonumber\\
&&\nonumber\\
&\, & \left.  +r^2 \cosh^2(2pt)  +\frac{ \left( dz+
pr^2d\varphi\right)^2}{\cosh(2pt) } \right] \,,\label{40} 
\end{eqnarray} 
where the irrelevant constant $1/\phi_0$ can be absorbed by a redefinition 
of units. 
(This mapping could have identified the Jordan frame scalar field as 
$\phi=\phi_0 \, \mbox{e}^{-ht}$.) So, Pimentel could have used the 
conformal map relating Jordan and Einstein frame instead of solving the 
Brans-Dicke field equations directly in \cite{Pimentel:1996du}.

Since $\phi>0$ implies $\phi_0>0$, in order to keep $\nabla^{\mu}\phi$ 
future-oriented with $\phi(t)=\phi_0 \, \mbox{e}^{-ht}$ we need $h>0$, 
hence \be\label{41} h=+\sqrt{ \frac{4(s-p^2)}{2\omega+3}} \,. \ee

\subsection{$\omega\to + \infty$ limit}

The Pimentel solution is one of those with anomalous limit to GR. In 
Brans-Dicke gravity, the standard lore is that the $\omega\to\infty$ limit 
of a solution reproduces the corresponding GR limit solution with the same 
matter source (e.g., \cite{Weinberg:1972kfs}).  Furthermore, the 
Brans-Dicke scalar is supposed to have the asymptotics \be \phi= 
\mbox{const.}+\mbox{O}\left( \frac{1}{\omega}\right) \label{42} \ee in the 
limit $\omega\to \infty$ \cite{Weinberg:1972kfs}. This is not true for 
electrovacuum solutions of Brans-Dicke gravity, which do not reduce to the 
corresponding GR solutions with the same matter and instead exhibit the 
asymptotics 
\be 
\phi= \mbox{const.}+\mbox{O}\left( 
\frac{1}{\sqrt{\omega}}\right) \,.\label{43} 
\ee 
This anomaly in the GR 
limit was initially reported for specific exact solutions 
\cite{Matsuda:1972zp,Romero:1992xx,Romero:1992bu,Romero:1992ci, 
Paiva:1993bv,Paiva:1993qa,Scheel:1994yn, 
Anchordoqui:1997du,Miyazaki:2000ij,Brando:2018kic, Nguyen:2024jgy} and then 
for general electrovacuum Brans-Dicke theory \cite{Banerjee:1996iy} and 
explained in \cite{Faraoni:1998yq,Faraoni:1999yp}.  The Pimentel solution  
falls into this category. In fact, when $\omega\to +\infty$, 
Eq.~(\ref{33}) gives 
\be 
h = \sqrt{ \frac{4(s-p^2)}{2\omega+3}} = 
\mbox{O}\left( \frac{1}{\sqrt{\omega}}\right)  \label{44} 
\ee 
and 
\be 
\phi(t) = \phi_0 \, \mbox{e}^{-ht} =\phi_0 +\mbox{O}\left( 
\frac{1}{\sqrt{\omega}}\right) \,. \label{45} 
\ee 
The straightforward 
limit of Pimentel's line element for $\omega\to +\infty$ (and, 
consequently, $\phi\to \phi_0$) is the Mars line element, which is {\em 
not} a vacuum solution. It is sourced by a stiff fluid or, alternatively, 
by a free scalar field which is nothing but the Einstein frame scalar 
field $\tilde{\phi}(t) = \tilde{\phi}_0 \, t$ of the Pimentel solution 
that should disappear in the limit, but does not. This behaviour is a 
manifestation of the general phenomenology of
 electrovacuum solutions of Brans-Dicke gravity, in which the limiting 
metric $g_{\mu\nu}^{(\infty)} $ of GR solves the Einstein equations with a 
``surviving'' scalar field as the source, i.e., the Einstein frame scalar 
field which does not disappear (see 
Refs.~\cite{Faraoni:2019sxw,Faraoni:2018nql} for a full 
discussion).\footnote{This conclusion is  reached without any reference 
to conformal transformations and the ``surviving'' scalar field is 
identified with the Einstein frame one only {\em a posteriori}.}

The limit $\omega\to +\infty$ gives $ h= \sqrt{ \frac{4(s-p^2)}{2 
\omega+3}} \to 0$ and $\phi(t)=\phi_0 \, \mbox{e}^{-ht} \to $~const. while 
  $s-p^2=0$ which, according to Eq.~(\ref{5}), produces 
  $\tilde{P}=\tilde{\rho}=0$. The Mars line element is reproduced but its 
  matter source is the new field $\tilde{\phi}(t) = -\sqrt{2(s-p^2)} \, 
  t$, not vacuum.

The thermal formalism of scalar-tensor gravity does not distinguish 
between GR and the anomalous limit of electrovacuum Brans-Dicke theory, 
which appear as two states of thermal equilibrium when using the Jordan 
frame \cite{Gallerani:2025myd}. However, the Einstein frame formulation of 
the formalism using chemical potential instead of temperature can 
distinguish between these two equilibria \cite{Gallerani:2025myd}. The 
study of this alternative formalism is beyond the scope of this article.

\subsection{New 3-parameter family of solutions}

The (Jordan frame) action of electrovacuum Brans-Dicke gravity with $T=0$, 
and the corresponding field equations, are invariant under the 
transformation \cite{Faraoni:1998yq,Faraoni:1999yp} \begin{eqnarray} 
g_{\mu\nu} & \to & \tilde{g}_{\mu\nu} = \phi^{2\alpha} g_{\mu\nu} \,, 
\label{symm1}\\
&&\nonumber\\
 \phi & \to & \bar{\phi} = \phi^{1-2\alpha} \quad\quad \left( \alpha\neq 
\frac{1}{2} \right) \,,\label{symm2}\\
&&\nonumber\\
\omega &\to & \tilde{\omega} = 
\frac{\omega+6\alpha(1-\alpha)}{(1-2\alpha)^2} \,, \label{symm3}\\
&&\nonumber\\
V(\phi) &\to & \bar{V}( \bar{\phi} ) = \bar{\phi}^{ 
-\frac{4\alpha}{1-2\alpha} } V\left( \bar{\phi}^{ \frac{1}{1-2\alpha} } 
\right) \label{symm4} 
\end{eqnarray} 
(this transformation has nothing to 
do with the transformation from Jordan to Einstein frame). This symmetry 
can now be used to generate a 3-parameter family of new solutions of 
electrovacum Brans-Dicke gravity by applying it to the Pimentel geometry. 
The result is 
\be 
\bar{\phi}= \phi^{ \frac{1}{1-2\alpha} }= \phi_0^{ 
\frac{1}{1-2\alpha} } \left( \mbox{e}^{-ht}\right)^{1-2\alpha} \equiv 
\bar{\phi}_0 \, \mbox{e}^{-\bar{h} t} \,, 
\ee 
where 
\be 
\bar{h}=h\left( 
1-2\alpha \right) = 2\left( 1-2\alpha \right )\sqrt{ 
\frac{s-p^2}{2\omega+3}} 
\ee 
and $\bar{\phi}_0 = \phi_0^{1-2\alpha} $. To 
keep $ \tilde{\nabla}^{\mu}\bar{\phi}$ future-oriented, it is necessary to 
choose $\alpha<1/2$ (but it is not necessary, of course, to impose this 
condition if one is not interested in applying the thermal formalism).  
The new line element is 
\begin{eqnarray} 
d\bar{s}^2 &=& \mbox{e}^{2\alpha 
ht} \left[ \mbox{e}^{sr^2} \cosh(2pt) \left( -dt^2+dr^2\right)
 \right.\nonumber\\
&&\nonumber\\
&\, & \left. + r^2
\cosh(2pt) d\varphi^2 + \frac{ \left( dz+ pr^2 
d\varphi\right)^2}{\cosh(2pt) } \right] 
\end{eqnarray} 
(dropping an 
irrelevant overall multiplicative constant $\phi_0^{1-2\alpha}$), while $ 
\tilde{\omega}$ is given by Eq.~(\ref{symm3}) and $\bar{V}( 
\bar{\phi})=0$. The properties of these new geometries are similar to 
those of the Pimentel spacetime and the causal structure is the same.

\subsection{Expansion scalar}

The expansion scalar $\Theta$ is important in order to check the ideas of
 the thermal analogy and test the results of 
Refs.~\cite{Faraoni:2018qdr,Faraoni:2021lfc,Faraoni:2021jri, 
Faraoni:2023hwu,Giardino:2023sw,Faraoni:2025alq}
 with a rather sophisticated example, which we do in the next sections.  
The four-velocity of the effective $\phi$-fluid associated with the 
Pimentel solution \begin{equation}
    u^\mu = \frac{\nabla^\mu \phi}{\sqrt{-\nabla^\alpha \phi \nabla_\alpha 
\phi}} = {\delta^\mu}_0 \, \frac{ \mbox{e}^{-sr^2/2} \, \mbox{e} 
^{-ht/2}}{\sqrt{\cosh \left( 2pt \right)}} \label{46bis} \end{equation} 
and the metric determinant with \begin{equation}
    \sqrt{-g} = \, \mbox{e}^{2ht} \, \mbox{e}^{sr^2} r \cosh\left( 2pt 
\right) \, , \label{48} \end{equation} give the expansion scalar 
\begin{eqnarray}
    \Theta &=& \frac{1}{\sqrt{-g}} \, \partial_\mu \left( \sqrt{-g} \, 
u^\mu \right)  \nonumber\\
&&\nonumber\\
&=&
\frac{ \mbox{e}^{-ht/2} \, \mbox{e}^{-sr^2/2}}{\cosh^{3/2}(2pt)}\left[ 
\frac{3h}{2} \cosh(2pt) + p \sinh(2pt) \right] \nonumber\\
&& \label{49}\\
& \equiv & A \left[ \frac{3h}{2} \cosh(2pt) + p \sinh(2pt) \right]
\label{expansionfactor} \end{eqnarray} where $A(t,r; h,p ) \equiv \frac{ 
\mbox{e}^{-\left( ht + sr^2 \right)/2} }{ \cosh^{3/2} \left( 2pt \right) } 
> 0$ for all values of the variables and of the parameters. We recover 
Mars' expansion scalar~(\ref{9}) in the limit in which $h \to 0$ (or $s 
\to p^2$).

In order to probe basic ideas of the thermal view of scalar-tensor 
gravity, we need to know the sign of the expansion $\Theta$ for all 
regimes of the Pimentel solution. The results are summarized in 
Table~\ref{table:1} (which also contains all the possible behaviours of 
${\cal K}{\cal T}$ derived in Sec.~\ref{sec:4}).

We rewrite the expansion scalar as \begin{equation}
    \Theta = \frac{A}{2} \, \mbox{e}^{-2pt} \left[ \mbox{e}^{4pt} \left( 
\frac{3h}{2} + p \right) +\frac{3h}{2} - p \right] \,, \end{equation} 
therefore, \be \Theta > 0 \quad \Leftrightarrow \quad \left[ 
\mbox{e}^{4pt} \left( \frac{3h}{2} + p \right) + \frac{3h}{2} -p \right] 
>0 \,. \ee We discuss separately the possible ranges of the parameters $p$ 
and $h$ in relation with each other.

\subsubsection{$p>0$}

For positive $p$, it is easy to see that if $p \leq 3h/2$, the expansion 
is positive for all times. However, if $p> 3h/2 $, then the sign of 
$\Theta$ depends on the value of $t$: \begin{align}
    \Theta > 0 \; \Leftrightarrow \; \mbox{e}^{4pt} > \frac{ p - 3h/2 }{ 
p+ 3h/2 } \; \Leftrightarrow \;
t>t_1
\end{align} 
where $p+ 3h/2 >0$ and 
\begin{equation}
    t_1 \equiv \frac{1}{4p} \, \ln \left( \frac{p - 3h/2}{p + 3h/2} 
\right) <0 \,. 
\end{equation} 
$t_1$ is negative since $0<p-3h/2 < p + 
3h/2$.

\subsubsection{$p<0$}

Now for negative $p$, if $-3h / 2 < p < 0$ the expansion $\Theta$ is 
positive at all times. However, if $p< -3h/ 2 < 0$ we have \begin{align}
    \Theta > 0 & \Leftrightarrow \frac{3h}{2} - p > -\mbox{e}^{4pt} \left( 
\frac{3h}{2}+p \right) \\
     & \Leftrightarrow \ln \left( \frac{p -3h/2}{p+ 3h/2} \right) >-4
|p| t \,.
\end{align} The left-hand side of this inequality is positive if the term 
inside the logarithm is larger than $1$; this is the case if $h>0$, which 
is always satisfied and this left-hand side is always positive. Now, 
\begin{equation}
    -4 |p| t < \ln \left( \frac{p-3h/2}{p+ 3h/2} \right) \quad \forall \, 
     t\geq 0 \,. \end{equation}

For negative $t$ we have \begin{align}
    \Theta > 0 & \Leftrightarrow t > -\frac{1}{4|p|}\ln \left( \frac{p 
-3h/2}{p+ 3h/2} \right)\\
     & \Leftrightarrow t > -t_2 \,,
\end{align} where 
\begin{equation}
    t_2 \equiv \frac{1}{4|p|} \, \ln \left( \frac{p-3h/2}{p+3h/2} \right) 
= \frac{1}{4|p|} \, \ln \left( \frac{|p| + 3h/2 }{|p| -3h/2} \right)
>0 \,.  \label{58}
\end{equation} 
To recap, we have \begin{align}
    t< -t_2 < 0 \quad & \Leftrightarrow \quad \Theta < 0 \, ,\\ 
\nonumber\\
    -t_2 < t < 0 \quad & \Leftrightarrow \quad \Theta > 0 \, ,\\ 
\nonumber\\
    t \geq 0 \quad & \Leftrightarrow \quad \Theta > 0 \, . \end{align}

\subsubsection{Range of $t$}

The expansion scalar $\Theta(t) $ is positive for all positive times $t$: 
the only term that could {\em a priori} become negative in the 
expression~(\ref{expansionfactor}) of the expansion $\Theta$ is the one 
containing the hyperbolic sine but this is $ p\sinh \left( 2pt \right)$, 
which is positive for all $t>0$ and for all $p\neq 0$, guaranteeing that 
$\Theta>0$ for positive times.
 
In the late-time limit $t \to +\infty$, the expansion scalar asymptotes to 
\begin{equation} \Theta \approx \sqrt{2} \, \mbox{e}^{-sr^2/2} \, 
\mbox{e}^{ -\left( |p| +h/2 \right) t} \left( \frac{3h}{2}+|p| \right)  
\to 0^+ \quad \, \forall \, p \neq 0 \end{equation} as $t\to+\infty$.

Now let us consider negative times $t<0$, for which $ p\sinh 
\left(2pt\right)<0$ for all $p\neq 0$. In the infinite past $t \to 
-\infty$, the asymptotic value of the expansion $\Theta$ depends on the 
value of $p$.

For positive $p$, the expansion scalar becomes \begin{align}
    \Theta &\approx \sqrt{2} \, \mbox{e}^{-ht/2} \, \mbox{e}^{-sr^2/2} \, 
\mbox{e}^{-3pt} \, \mbox{e}^{-2pt} \left( \frac{3h}{2} - p \right) \\
    &= \sqrt{2}\, \mbox{e}^{-sr^2/2}\left( \frac{3h}{2} - p \right)
\mbox{e}^{\left( h/2 + 5p \right) |t| } \label{limit-t+p} \,. \end{align} 
$\Theta$ diverges for all positive values of $p\neq 3h/2$, but its sign 
depends on the exact values of $h$ and $p$. For $ 0<p< 3h/2 $, $\Theta$ 
remains positive. However, for $0< 3h/2 <p $, the limit becomes negative 
and for $p=3h/2$ the expansion scalar $\Theta \to 0$ as $t\to -\infty$.

For negative $p$, in the limit $t \to -\infty$ the expansion scalar 
becomes \begin{align}
    \Theta &\approx \sqrt{2}\, \mbox{e}^{-ht/2} \, \mbox{e}^{-sr^2/2} \, 
\mbox{e}^{-3p t} \, \mbox{e} ^{2pt} \left( \frac{3h}{2} + p \right)\\
    & = \sqrt{2}\, \mbox{e}^{-sr^2/2} \left( \frac{3h}{2} - |p|
\right)  \mbox{e}^{ \left( h/2 - |p| \right) |t|} \,. \end{align} 
Therefore, we have a similar result as in Eq.~(\ref{limit-t+p}) so the 
expansion scalar behaves as follows in the limit $t \to -\infty$: 
\begin{align}
     \text{if} \;\;\; 0<|p|< \frac{h}{2} \,, \quad \quad & \quad \lim_{t 
\to -\infty} \Theta = +\infty \,;\\
     \text{if} \;\;\; 0< \frac{h}{2}<|p| \,, \quad\quad & \quad \lim_{t 
\to -\infty} \Theta = 0 \,,\\
 \text{if} \;\;\; p = \frac{h}{2} \,, \quad \quad & \quad \lim_{t \to 
-\infty} \Theta = \sqrt{2}\, h \mbox{e}^{-sr^2/2} \,,\\
 \text{if} \;\;\; p = -\frac{3h}{2} \,, \quad \quad & \quad \lim_{t \to 
-\infty} \Theta = 0 \,. \end{align}

We now proceed to analyze the implications of the behaviour of $\Theta$ 
for the thermal approach to scalar-tensor gravity.

\small \begin{widetext} \begin{center} \begin{table}
    \begin{tabular}{| c | c | c | c | c |} \hline
& $t>0$ & $t \to + \infty$ & $t<0$ & $t\to-\infty$\\
\hline \multirow{2}{1.2em}[-2em]{$\Theta$} & 
\multirow{2}{4em}[-2em]{$\Theta >0$}
& \multirow{2}{4.9em}[-2em]{\makecell{$\Theta \to 0^+ $\\
$\forall \; p \neq 0$ }}& \makecell[l]{$p<0:$ \\ $\Theta >0 \quad 
\text{if} \quad - 3h/2 <p<0$ \\ $\Theta >0 \quad \text{if} \quad p< -3h/2 
\quad \text{and} \quad t>-t_2 $\\ $\Theta < 0 \quad \text{if} \quad 
p<-3h/2 \quad \text{and} \quad t<-t_2 $} & \multirow{2}{14em}[-1.3em]{ 
\makecell{$ \Theta \to +\infty \quad \text{if} \quad 0<|p|<h/2 $\\ $\Theta 
\to 0 \quad \text{if} \quad 0< h/2 < |p|$\\ $\Theta \to \sqrt{2}\, 
h\mbox{e}^{-sr^2} \quad \text{if} \quad |p|= h/2 $}}\\ \cline{4-4}& & & 
\makecell[l]{$p>0:$ \\ $\Theta >0 \quad \text{if} \quad p \leq 3h/2 $\\ 
$\Theta >0 \quad \text{if} \quad p> 3h/2 \quad \text{and} \quad t>t_1$\\ 
$\Theta <0 \quad \text{if} \quad p> 3h/2 \quad \text{and} \quad t<t_1$} 
&\\ \hline \multirow{2}{1.6em}[-2em]{${\cal K}{\cal T}$} & 
\multirow{2}{5em}[-2em]{\makecell{$ 8\pi {\cal K}{\cal T} <\Theta$\\ 
$\forall \; p \neq 0$}} & \multirow{2}{6.3em}[-0.75em]{\makecell{$ {\cal 
K}{\cal T} \to 0^+$\\ $\frac{ 8\pi {\cal K}{\cal T}}{\Theta} \sim 
\frac{h}{\frac{3h}{2}+|p|}$}}
&  
\makecell[l]{$p>0:$\\ $8\pi {\cal K}{\cal T} <\Theta \quad \text{if} \quad 
p< h/2$\\ $8\pi {\cal K}{\cal T} >\Theta \quad \text{if} \quad p>h/2 \quad 
\text{and} \quad t<t_3$\\ $8\pi {\cal K}{\cal T} <\Theta \quad \text{if} 
\quad p> h/2 \quad \text{and} \quad t > t_3$} & 
\multirow{2}{10em}[-1em]{\makecell[l]{${\cal K}{\cal T} \to +\infty \quad 
\text{if} \quad |p|< h/2 $\\ ${\cal K}{\cal T} \to {\cal J}_0 
\quad \text{if} \quad |p| = h/2 $\\ ${\cal K}{\cal T} \to 0^+ 
\quad \text{if} \quad |p|> h/2 $\\ $\frac{ 8\pi{\cal K}{\cal T}}{\Theta} 
\sim \frac{h}{\frac{3h}{2}-|p|}$}}\\ \cline{4-4}
&  &  & \makecell[l]{$p<0:$\\
$ 8\pi {\cal K}{\cal T} <\Theta \quad \text{if} \quad - h/2 \leq p < 0 $\\ 
$ 8\pi {\cal K}{\cal T} >\Theta \quad \text{if} \quad p< -h/2 \quad 
\text{and} \quad t<t_3$\\ $ 8\pi {\cal K}{\cal T} <\Theta \quad \text{if} 
\quad p<- h/2 \quad \text{and} \quad t_3<t<0$} & \\ \hline \end{tabular} 
\caption{Summary of the thermal view of Pimentel's solution of Brans-Dicke 
gravity. $t_2$, $t_3$, and ${\cal J}_0(r)$  are given by Eqs.~(\ref{58}),  
(\ref{t3}), and~(\ref{J0}), respectively.\label{table:1}} 
\end{table} 
\end{center} 
\end{widetext} \normalsize

\section{Thermal view of Pimentel spacetime} 
\label{sec:4} 
\setcounter{equation}{0}

In this section we switch to units in which $ G=1$ to facilitate 
comparison with the general thermal view of scalar-tensor gravity 
\cite{Faraoni:2018qdr,Faraoni:2021lfc,Faraoni:2021jri, 
Faraoni:2023hwu,Giardino:2023sw,Faraoni:2025alq}. It has been shown in 
Ref.~\cite{Miranda:2022wkz} that, for 
``first-generation'' scalar-tensor gravity, including Brans-Dicke theory, 
the viscous stresses of the effective $\phi$-fluid depend only on the 
first derivatives of the four-velocity, and this effective fluid behaves 
as a Newtonian fluid.  The constitutive relations $ P_\mathrm{viscous} = 
-\eta \, \Theta $ and $\pi_{\mu\nu}=-2\zeta \, \sigma_{\mu\nu}$ are 
satisfied \cite{Miranda:2022wkz}, where $\eta $ and $\zeta$ are bulk and 
shear viscosity coefficients, respectively. This result applies to the 
effective fluid of the Pimentel solution as well.  More general 
scalar-tensor theories, such as Horndeski theories (even so-called 
``viable'' ones \cite{Langloisetal18} satisfying the astrophysical 
constraints imposed by the equality between the speed of gravitational 
waves and the speed of light 
\cite{LIGOScientific:2017vwq,LIGOScientific:2017zic}) do not satisfy these 
constituive relations, and give rise to non-Newtonian effective fluids 
\cite{Miranda:2022wkz}.

\subsection{Thermal history}

For the Pimentel spacetime we have 
\be 
{\cal K}{\cal T}= \frac{ 
\sqrt{-\nabla^{\alpha} \phi \nabla_{\alpha} \phi}}{8\pi \phi} = \frac{ h\, 
\mbox{e}^{-ht/2} \, \mbox{e}^{-sr^2/2}}{ 8\pi \sqrt{ \cosh \left( 2pt 
\right)} } \,.
\ee 
By definition, ${\cal K}{\cal T} >0$ for all $t$ and all 
values of $p$. When $\Theta >0$, the critical half-line $8\pi {\cal 
K}{\cal T}=\Theta \geq 0 $ is given by 
\begin{equation}
 \frac{ 8\pi {\cal K}{\cal T}}{\Theta} = \frac{h}{ p \tanh (2pt) + 3h/2 
} \to \frac{2h}{3h \pm 2p} \quad \mbox{as} \; t\to \pm \infty \,. 
\end{equation} 
In the special (and fine-tuned) case $p=-h/2$, the critical 
half-line $8\pi {\cal K}{\cal T}=\Theta$ is approached as $t\to +\infty$.

As already remarked, the Pimentel spacetime does not contain 
singularities, which originated interest in it and in previous solutions 
\cite{Senovilla:1990rt,Chinea:1992nq,Ruiz:1992np, 
Mars:1995jv,Pimentel:1996du}. Similarly, the scalar $\phi=\phi_0\, 
\mbox{e}^{-ht}$ cannot vanish, therefore we do not expect infinite 
deviations from GR (i.e., divergences in ${\cal K}{\cal T}$).  However, 
gravity becomes stronger as $r\to0$ and indeed, as a function of $r$, 
${\cal K}{\cal T}$ is maximum on the $z$-axis, validating one of the basic 
ideas of the thermal approach to scalar-tensor gravity, that gravity 
deviates more from GR where it is stronger. However, the deviation from GR 
does not necessarily go with the focusing of the fluid lines of the 
effective fluid, corresponding to expansion scalar $\Theta<0$, as 
hypothesized in \cite{Faraoni:2021lfc,Faraoni:2021jri}. If fact, the 
absolute value of $\Theta$ is larger at $r=0$, but its sign varies with 
the parameters $h,p$ and with time. Hence, the deviation of gravity from 
GR should be assessed using ${\cal K}{\cal T}$ or the strength of gravity, 
not the sign of $\Theta$. It is true, however, that $\Theta<0$ ``heats 
up'' gravity because then the term $-8\pi {\cal K}{\cal T} \Theta$ in the 
right hand side of Eq.~(\ref{evolution-reduced}) contributes to making 
$d({\cal K}{\cal T})/d\tau$ positive. In the limit $t \to +\infty$, 
\begin{equation}
    {\cal K}{\cal T} \propto \mbox{e}^{- \left( h/2 + |p| \right)t} 
\end{equation} and, since $h>0$, $ |p|+h/2 >0$ and $ \lim_{t \to +\infty} 
{\cal K}{\cal T} = 0^+ $.

In the limit $t \to -\infty$, 
\begin{equation}
    {\cal K}{\cal T} \propto \, \mbox{e}^{- \left( h/2 - |p| \right)t} = 
\, \mbox{e}^{ \left( h/2 - |p| \right) |t|} 
\end{equation} 
and 
\begin{align}
    \lim_{t \to -\infty} {\cal K}{\cal T} = +\infty \quad &\Leftrightarrow 
\quad |p| < \frac{h}{2} \,,\\
    \lim_{t \to -\infty} {\cal K}{\cal T} = {\cal J}_0(r) \equiv 
\frac{ \sqrt{2}\, h \, \mbox{e}^{-sr^2/2}}{8\pi} \quad
&\Leftrightarrow
\quad |p| = \frac{h}{2} \,,\\
    \lim_{t \to -\infty} {\cal K}{\cal T} = 0^+ \quad &\Leftrightarrow 
\quad |p| > \frac{h}{2} \,.\label{J0}
\end{align}

\subsubsection{Relation between comoving and coordinate times}

The proper time $\tau$ of the effective $\phi$-fluid and the coordinate 
time $t$ do not coincide. In fact, 
\be 
u^0=\frac{dt}{d\tau} =\frac{ 
\mbox{e}^{-sr^2/2} \, \mbox{e}^{-ht/2} }{\sqrt{ \cosh(2pt)}} >0\,, 
\ee 
so 
$\tau$ increases when $t$ increases and {\em vice-versa}. We have 
\be 
\tau = \frac{ \mbox{e}^{sr^2/2} }{\sqrt{2}} \, \int dt \, \sqrt{ 
\mbox{e}^{ ( h+2p) t} \,+ \, \mbox{e}^{( h-2p) t} } \,, 
\ee 
which can only be integrated explicitly using an hypergeometric function 
$_2F_1$, 
\begin{eqnarray} 
\tau &=& 
  -\sqrt{2} \, \mbox{e}^{sr^2/2} \, \frac{\left( \mbox{e}^{4pt} 
  +1\right)}{ 2p-h} \, \sqrt{ \left( \mbox{e}^{ 4pt} \, + 1 \right) \, 
  \mbox{e}^{( h-2p) t} } \nonumber\\
&\, & \times \, _2F_1\left( 1, \frac{h}{8p}+\frac{5}{4} ;
\frac{h+6p}{8p}; -\mbox{e}^{4pt} \right) \,, 
\end{eqnarray} 
but this is not useful. Let us look instead at the limits $t\to \pm 
\infty$.

As $ t\to +\infty$, 
\be 
d\tau \simeq \frac{ \, \mbox{e}^{sr^2/2} }{\sqrt{2}} \, 
\mbox{e}^{ \left( |p| + h/2 \right) t} dt 
\ee 
and 
\be 
\tau \simeq \frac{ 
\mbox{e}^{sr^2/2} }{\sqrt{2} \left( |p| +h/2 \right)} \, \mbox{e}^{ \left( 
|p| + h/2 \right) t} \,, 
\ee 
hence $\tau \to +\infty$ as $ t\to +\infty$.

In the other limit $t\to -\infty$, we obtain 
\be 
d\tau \simeq \frac{ 
\mbox{e}^{sr^2/2}}{\sqrt{2} } \, \mbox{e}^{ \left( -|p| + h/2 \right) t} 
dt 
\ee 
and 
\be 
\tau \simeq  \frac{ \, \mbox{e}^{sr^2/2} }{ \sqrt{2} 
\left( -|p| +h/2 \right) } \, \mbox{e}^{ \left( -|p| + h/2 \right) t } \,. 
\ee 
Then, if $|p|<h/2$, it is $\tau>0$ and $\tau\to 0^{+}$ when $t\to 
-\infty$ (that is, $\tau$ begins from zero). If instead $|p|>h/2$, it is 
$\tau<0$ and $\tau\to -\infty$ as 
$t\to -\infty$.

The thermal behaviour of the Pimentel geometry as the location in the 
$\left( h,p \right)$ parameter space changes can be analyzed using 
Table~\ref{table:1}.\\

\subsubsection{$t<0$ and $p < -3h/2$}

Evolution in the $\left( \Theta, {\cal K}{\cal T}\right)$ plane is 
significantly more involved for $t<0$ than for $t>0$, due to the more 
varied range of possibilities arising in the $\left(h,p \right)$ parameter 
space, as is clear from Table~\ref{table:1}.

A straightforward analysis for $p<-3h/2$ shows that $ \Theta <0$ for $t <- 
t_2$ and $\Theta>0$ for $ -t_2 <t<0$, while $ 8\pi {\cal K}{\cal T}
>\Theta
$ if $t<t_3<0$ and $ 8\pi {\cal K}{\cal T} < \Theta $ for $t_3 <t <0$, 
where $t_2$is given by Eq.~(\ref{58}) and 
\be
t_3 \equiv - \frac{1}{2p}\text{arctanh}\left( \frac{h}{2p} \right)  
\label{t3}
\ee 
is obtained by setting $8\pi {\cal K}{\cal T}=\Theta$.  

It is 
important to compare $t_2$ and $t_3$. 
Appendix~\ref{Appendix:A} shows that it is always $-t_2< t_3$.  Using 
Table~\ref{table:1}, we can state that:

\begin{itemize}

\item If $t<-t_2$, then $\Theta<0$ and gravity ``heats up'';

\item If $ -t_2 < t < t_3$, then $\Theta>0$ and $8\pi {\cal K}{\cal T}
>\Theta$; gravity ``heats up'' again;

\item If $ t_3 <t <0$, it is $\Theta>0$ and $8\pi {\cal K}{\cal T}<\Theta 
$, hence gravity ``cools''.

\end{itemize}

The trajectory of the Pimentel 
spacetime in the $\left( \Theta, {\cal K}{\cal T} \right)$ plane  
crosses the critical half-line $8\pi {\cal K}{\cal T}=\Theta > 0$ 
with horizontal tangent since $ d\left( {\cal K}{\cal T} 
\right)/dt \to 0$ as $t\to t_3$ (but $d\Theta/dt$ does not vanish).  
Simultaneously, $ d\left( {\cal K}{\cal 
T} \right)/d\Theta \to 0$ and this trajectory becomes instantaneously  
horizontal. In fact, 
\begin{eqnarray} 
\frac{d\Theta}{dt} &=& \frac{ \, 
\mbox{e}^{-sr^2/2} \, \mbox{e}^{-ht/2} }{\cosh^{3/2} \left( 2pt \right)} 
\left\{-\left[ \frac{h}{2}
+3p\tanh(2pt)\right] \right.\nonumber\\
&&\nonumber\\
&\, & \left.
\times \left[ \frac{3h}{2} \, \cosh(2pt) +p \sinh(2pt)\right] 
\right.\nonumber\\
&&\nonumber\\
&\, & \left.
+ 2 p\left[ \frac{3h}{2} \, \sinh(2pt) +p\cosh(2pt) \right]\right\} \,, 
    \end{eqnarray}

\be 
\frac{ d\left( {\cal K}{\cal T} \right)}{dt} = -\, \frac{h \, 
\mbox{e}^{-sr^2/2} \, \mbox{e}^{-ht/2} }{16 \pi \cosh^{3/2} (2pt) } \left[ 
h\cosh(2pt)+2p\sinh(2pt) \right] \,. 
\ee
 At $t_3$, which is defined by $h\cosh(2pt_3)+ 2p\sinh(2pt_3)=0$, $d\left( 
{\cal K}{\cal T}\right)/dt $ vanishes, while
\be
\frac{d\Theta}{dt} \Big|_{t_3} = 
\frac{ \mbox{e}^{-sr^2/2} \, \mbox{e}^{-ht/2} }{ \cosh^{3/2} \left( 
2pt_3 \right) } 
\left( 3h-\frac{4p^3}{h} -4ph \right) \sinh\left( 2pt_3 \right) 
\neq 0 \,.
\ee
We also have 
\begin{widetext} 
\begin{eqnarray} 
\frac{ d\left( {\cal K}{\cal T}\right)}{d\Theta} &=& 
\frac{ d\left( {\cal K}{\cal T}\right)}{dt} \, \frac{ dt}{d\Theta} 
\nonumber\\
&&\nonumber\\
&=& \frac{ -h \left[ h \cosh(2pt)+2p\sinh(2pt) \right]}{16\pi \left\{
-\left[ \frac{h}{2} +3p\tanh(2pt)\right] \left[ \frac{3h}{2} \, \cosh(2pt) 
+p\sinh(2pt) \right]
+2p \left[ \frac{3h}{2} \, \sinh(2pt) +p \cosh(2pt) \right] \right\} } \,, 
 \end{eqnarray} 
\end{widetext} 
therefore, also $d\left( {\cal K}{\cal 
 T}\right)/d\Theta =0$ at $t=t_3$.

At $t_3$,  
\be 
\Theta(t_3) =8\pi {\cal K}{\cal 
T}(t_3)  = \frac{ \sqrt{2} \, h \, \mbox{e}^{-sr^2/2} }{\sqrt{ 
\mbox{e}^{(h+2p)t_3 } 
+ \, \mbox{e}^{ (h-2p)t_3} } } \,. 
\ee 
The trajectory of the Pimentel 
spacetime  crosses the critical half-line $8\pi {\cal K}{\cal 
T}=\Theta \geq 0$, as illustrated in Fig.~\ref{fig:pheno1} where ${\cal 
K}{\cal T}(\Theta)$ is plotted parametrically using $t$ as the parameter. 
On the other side of the critical half-line (i.e., for $t>t_3$), 
the trajectory continues below this half-line, with $8\pi {\cal K}{\cal 
T}<\Theta$, and  approaches GR: gravity ``cools'' there. This is new 
phenomenology not seen before in any exact solution of Brans-Dicke gravity 
(or of other scalar-tensor gravities) and not contemplated in the general 
theory \cite{Faraoni:2018qdr,Faraoni:2021lfc,Faraoni:2021jri, 
Faraoni:2023hwu,Giardino:2023sw,Faraoni:2025alq,Faraoni:2025alq}.

\begin{figure}
    \centering \includegraphics[width=0.85\linewidth]{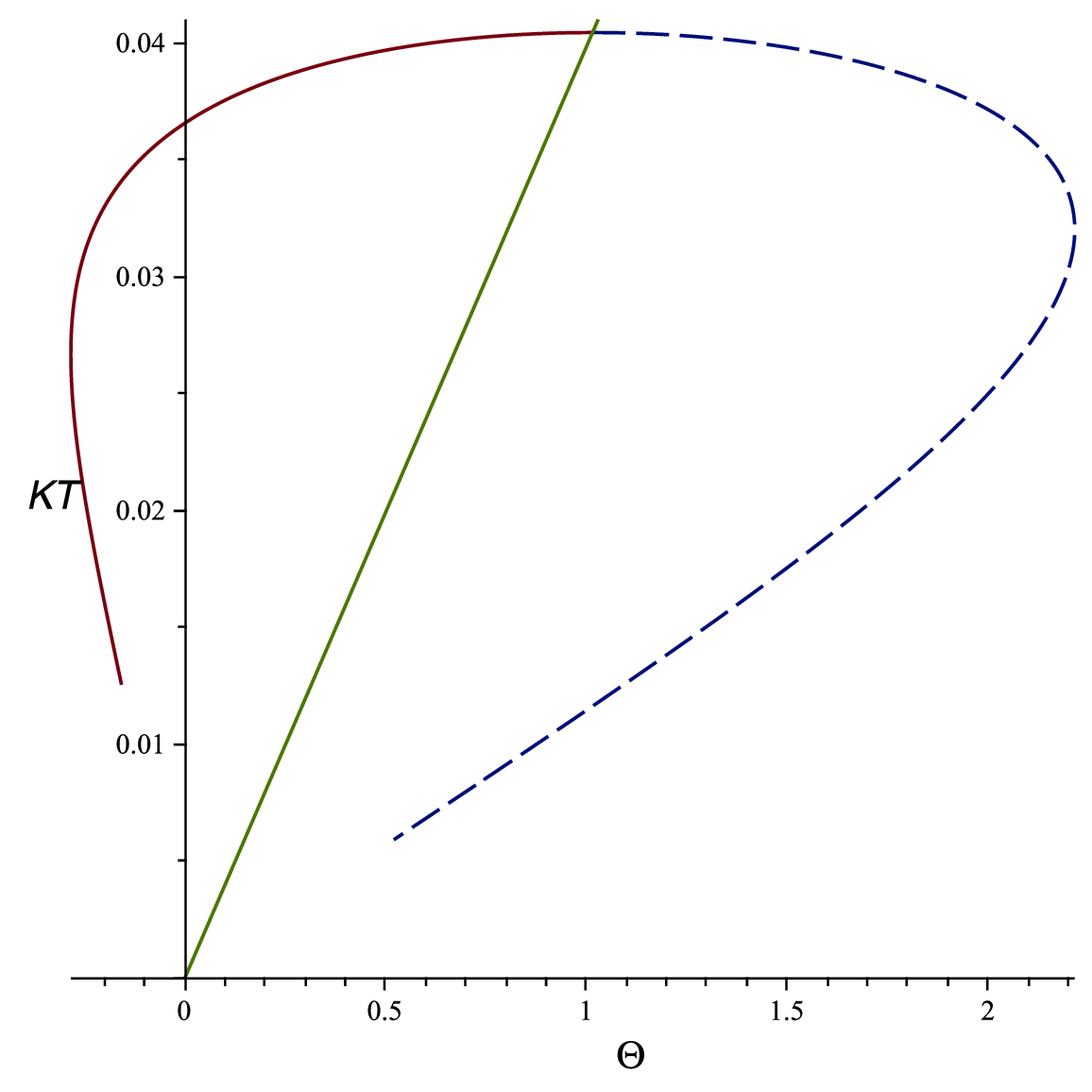} 
    \caption{In the $\left( \Theta, {\cal K}{\cal T}\right)$ plane, the
trajectory of the Pimentel spacetime (solid)  begins in the $\Theta<0$ 
region, crosses the vertical axis and continues in the region $\Theta>0$, 
$8\pi {\cal K}{\cal T}>\Theta$, where it approaches  
the critical half-line $8\pi {\cal K}{\cal T}=\Theta$, crossing it with 
horizontal tangent. For $t>t_3$, the trajectory continues (dashed) in the 
region below the critical 
half-line. (For illustration, $s=0$, $p=-2h$, the time parameter $t$ is in 
units 
of $h^{-1}$, while $\Theta$ and ${\cal K}{\cal T}$ are in units of $h$ and 
different scales are used on the axes for clarity.)}\label{fig:pheno1} 
\end{figure}

\subsubsection{$t<0$ and $-3h/2 < p < -h/2 $}

In this case it is always $\Theta>0$, while 
\be 
8\pi {\cal K}{\cal T} 
\left\{ \begin{array}{cc}
> \Theta & \mbox{if} \quad t<t_3 \,,\\
&\\
< \Theta & \mbox{if} \quad t_3<t<0 \,, \end{array}\right. 
\ee 
hence 
gravity ``heats up'' for $t<t_3$. At $t=t_3<0$, there is a point on 
the half-line $8\pi {\cal K}{\cal T}=\Theta$, which is approached 
with horizontal tangent  as $t\to t_3^{-}$. After crossing the critical 
half-line at this point, gravity 
``cools'' since $\Theta $ becomes positive and $ 8\pi {\cal K}{\cal 
T}<\Theta$.

\subsubsection{$t<0$ and $ -h/2< p < 0$ or $ 0<p<h/2$}

In this case $\Theta >0$ and $8\pi {\cal K}{\cal T} <\Theta $, hence 
gravity ``cools''. Indeed, both derivatives 
\be 
\frac{d \left( {\cal 
K}{\cal T} \right)}{dt} = -\, \frac{h \, \mbox{e}^{-sr^2/2} \, 
\mbox{e}^{-ht/2} }{16\pi \sqrt{ \cosh (2pt)}} \left[ h + 2p \tanh( 2|p|t) 
\right] 
\ee 
and 
\be 
\frac{d \left( {\cal K}{\cal T} \right)}{d\tau} = -\, 
\frac{h \, \mbox{e}^{-sr^2} \, \mbox{e}^{-ht} }{8\pi \cosh (2pt)} \, 
\left[ \frac{h}{2} +|p| \tanh( 2|p|t) \right] 
\ee 
are negative.

\subsubsection{$t<0$ and $ h/2< p < 3h/2$}

In this case, $\Theta>0$ and $8\pi {\cal K}{\cal T}>\Theta $ if $t<t_3$: 
gravity ``heats up'' at these early times, while at $t=t_3$ the expansion 
$\Theta(t_3) >0$ and, as $t\to t_3^{-}$, $ d\left( {\cal K}{\cal 
T}\right)/dt \to 0 $ and $ d\left( {\cal K}{\cal T}\right)/d\Theta \to 0 
$. The point $\left( \Theta(t_3), \frac{ \Theta(t_3)}{8\pi} \right)$ 
is approached with horizontal tangent. At $t>t_3$, the trajectory  
continues in the region $8\pi {\cal K}{\cal T}<\Theta$, gravity ``cools'', 
and GR is approached.

\subsubsection{$t<0$ and $p > 3h/2$}

First, note that it is always $t_1<t_3$ (Appendix~\ref{Appendix:B}).

If $t<t_1$ it is $\Theta<0$ and gravity ``heats up''. At $t=t_1$, $\Theta 
$ vanishes and the vertical axis $\Theta=0$ is crossed at a value ${\cal 
K}{\cal T}(0)  >0$. Then, for $t_1 
<t< t_3$, it is $\Theta>0$ and $8\pi {\cal K}{\cal T}>\Theta>0$ and 
gravity 
continues to ``heat up'' until the time $t_3$. As $t_3$ is approached, the 
trajectory becomes horizontal and crosses the critical half-line at  
 $ \left( \Theta(t_3) , 
\frac{ \Theta(t_3)}{8\pi} \right)$, continuing at $t>t_3$ 
in the region $8\pi {\cal K}{\cal T}<\Theta$ below,  where gravity 
``cools'' and GR is approached.

\subsubsection{Thermal behaviour for $t>0$}

At $t=0$ we have \be \Theta(0) = \frac{3h}{2} \, \mbox{e}^{-sr^2/2} >0 \ee 
and $ds^2 \Big|_{0} = ds^2_\textrm{Mars} \Big|_0 $, so at $t=0 $ the 
Pimentel spacetime expands for all radii.

For $t>0$, the expansion scalar $\Theta $ is always positive, with 
$\Theta\to 0^{+}$ as $t\to +\infty$. The point representing the Pimentel 
spacetime in the $\left( \Theta, {\cal K}{\cal T}\right)$ plane moves 
toward decreasing values of $\Theta$.  Furthermore, $8\pi {\cal K}{\cal 
T}<\Theta$ for all values of $p\neq 0$, therefore, a state of thermal 
equilibrium is approached asymptotically, with ${\cal K}{\cal T}\to 0^{+}$ 
as $t\to +\infty$. This asymptotic future state does not look like GR 
since the Pimentel line element $ds^2 = \, \mbox{e}^{ht} \, 
ds^2_\textrm{Mars}$ still differs from the Mars one. This fact is not 
surprising since the Pimentel scalar field $\phi(t)=\phi_0 \, 
\mbox{e}^{-ht} \to 0$ as $t\to +\infty$: gravity becomes infinitely strong 
in this limit with the effective gravitational coupling $G_\mathrm{eff} 
\simeq 1/\phi \to +\infty$, which is certainly not a GR feature. The 
$\Theta\to 0^{+}$, ${\cal K}{\cal T} \to 0^{+}$ limit as $t \to +\infty $ 
is physically pathological, although it formally obeys the predictions 
obtained in 
the $\left( \Theta, {\cal K}{\cal T} \right)$ plane in 
\cite{Faraoni:2025alq}. The origin of this plane can 
be a physically pathological state.

\section{Conclusions} 
\label{sec:5} 
\setcounter{equation}{0}

Since the Mars and the Pimentel cylindrical geometries do not 
describe situations that we expect to encounter in nature, the interest in 
them is theoretical.  
The main purpose of this work consists of shedding new light on the 
thermal view of scalar-tensor gravity, using an appropriate family of 
exact solutions of Brans-Dicke gravity in which specific behaviours can be 
studied explicitly, as opposed to speculating about them, as  one is 
limited to do in the general theory. We have chosen the Pimentel geometry 
to do this.

The Pimentel solution has the advantage that, while being inhomogeneous, 
its scalar field $\phi(t)=\phi_0 \, \mbox{e}^{-ht}$ depends only on time, 
which makes its gradient $\nabla^{\mu} \phi$ timelike and amenable to 
treatment with the new thermal view. Moreover, it is a vacuum solution 
without scalar field potential, which simplifies the thermal physics since 
then $\Box \phi=0$ and the corresponding ``heat equation'' assumes the 
particularly simple form~(\ref{evolution-reduced})  studied in 
\cite{Faraoni:2025alq}. Even with these simplifications, we have uncovered 
new possibilities for the thermal behaviour of scalar-tensor gravity, 
which shows that the thermal analogy is richer than expected.

Before launching into the thermal analysis of the Pimentel 
geometry, we have studied its properties and its relation to the Mars 
solution of GR. This study has some value in itself, as these aspects were 
not studied before (Ref.~\cite{Pimentel:1996du} presents the Pimentel 
geometry without discussing its properties). First, we have shown that the 
Mars solution of GR, originally introduced with a vorticity-free stiff 
fluid as a source, can also be sourced by a scalar field, which may be 
useful when one searches for examples of exact scalar field solutions of 
GR.  One would not normally think of looking for the corresponding scalar 
field source, but this is natural when attempting to relate the GR 
geometry to a scalar-tensor one, as we did here. Second, we have shown how 
the Pimentel geometry is nothing but the Jordan frame version of the 
(Einstein frame) Mars one in Brans-Dicke gravity, which could have lead to 
the discovery of the Pimentel solution almost without calculations. This 
result establishes a new connection between Pimentel and Mars geometries, 
which could be useful in the future to compare the thermal view of 
scalar-tensor gravity in the Jordan frame (which uses temperature at zero 
chemical potential) with the Einstein frame description (at zero 
temperature but non-vanishing chemical potential) 
\cite{Faraoni:2022gry,Gallerani:2025myd}. The former is under development, 
while the latter is almost completely unexplored. 

Finally, we have shown how the Mars solution sourced by a scalar field is 
the $\omega\to+\infty$ ``anomalous'' limit of the Pimentel solution 
(anomalous in the sense that the Brans-Dicke scalar field, which should 
vanish in this limit, instead survives becoming a matter source instead of 
a gravitational degree of freedom). Thus, we add another example to the 
catalogue of exact solutions of electrovacuum Brans-Dicke gravity with 
this property. (Again, one would not suspect the Mars solution to have 
this property if one did not attempt to relate it to the Pimentel 
geometry, as we do here.) The anomalous limit has the potential 
implication to invalidate future second order tests of gravity in the 
Solar System \cite{Faraoni:2019sxw}.

Moving on to the thermal view of scalar-tensor gravity, which is 
the most interesting part of this work,} the two-parameter Pimentel class 
of solutions of ``simple'' Brans-Dicke gravity offers new insight in the 
thermal view of scalar-tensor gravity.  The critical half-line $8\pi {\cal 
K}{\cal T}=\Theta \geq 0$ can be crossed dynamically, during the evolution 
of spacetime (Fig.~\ref{fig:pheno1}), but only with horizontal tangent $d 
\left( {\cal K}{\cal T}\right)/d\Theta = 0$. In the region $8\pi {\cal 
K}{\cal T} >\Theta$ gravity ``heats up'' running away from GR, while in 
the region $8\pi {\cal K}{\cal T}<\Theta$ gravity ``cools'' and GR is 
approached.  This is new phenomenology with respect to the general theory 
formulated in Refs.~\cite{Faraoni:2018qdr,Faraoni:2021lfc,Faraoni:2021jri, 
Faraoni:2023hwu,Giardino:2023sw,Faraoni:2025alq} and the known behaviours 
of exact solutions studied in \cite{Faraoni:2022doe, 
Faraoni:2022jyd,Giardino:2023qlu, Faraoni:2022fxo, Houle:2024sxs, 
Karolinski:2024nwp}.

It is indeed difficult to develop the thermal 
picture of scalar-tensor gravity by studying only the general theory (as 
done 
in Ref.~\cite{Faraoni:2025alq} that missed the critical line crossing). 
Exact solutions act as laboratories for identifying new phenomenology and 
new aspects of the thermal view of scalar-tensor gravity. As seen in the 
previous sections, even in relatively simple situations in which 
Eq.~(\ref{evolution_general2}) simplifies to the 
form~(\ref{evolution-reduced}) by removing effective sources or sinks of 
``heat'', it is far from trivial to extract results.

Overall, the value of the new thermal view of scalar-tensor gravity is 
that, in spite of its limits of applicability, it provides a coherent 
view describing phenomena appearing in areas as diverse as cosmology 
\cite{Giardino:2022sdv,Houle:2024sxs,Giardino:2022sdv,Faraoni:2025fjq}, 
black holes \cite{Faraoni:2025ufi},  exact solutions 
\cite{Faraoni:2022jyd,Giardino:2023qlu,Faraoni:2022fxo,Karolinski:2024nwp}, 
non-dynamical theories as anomalous  states of equilibrium 
\cite{Faraoni:2022doe},  limit to GR, etc. 
with few ideas. 
Basically, the approach to GR (which could have occurred in the early 
universe, at least on scales much smaller than the Hubble radius 
\cite{Damour:1992kf,Damour:1993id,Serna:2002fj,Faraoni:2025alq}) is akin 
to the relaxation of a hot fluid to its thermal equilibrium, while strong 
gravity regions act as ``hot spots'' 
\cite{Faraoni:2021lfc,Faraoni:2021jri}. These ideas allow to frame 
simultaneously many results obtained in the literature since the 1960s 
 in a single view. Certain results, including some now almost forgotten,  
could have been predicted based on the thermal description (see. e.g., 
\cite{Faraoni:2025fjq} for $f(R)$ cosmology). In the 
simplest situations (i.e., when $\Box\phi=0$), one generically expects 
gravity to deviate from GR behaviour near spacetime singularities, with 
implications that are only beginning to be understood. The extension to 
Horndeski gravity and the characterization of specific theories of gravity 
using the thermal and rheological properties of the effective $\phi$-fluid 
are just beginning to be studied \cite{Giusti:2021sku,Miranda:2022wkz}. 
Future work will continue the development of the thermal formalism.

\begin{acknowledgments}

V.~F. is supported, in part, by the Natural Sciences \& Engineering 
Research Council of Canada (Grant no. 2023-03234).

\end{acknowledgments}

\begin{appendices}

\section{Proof that $-t_2<t_3$} \label{Appendix:A} 
\renewcommand{\theequation}{A.\arabic{equation}} \setcounter{equation}{0}

The inequality $-t_2<t_3$ is equivalent to 
\be 
-\, \frac{1}{4|p|} \, \ln 
\left( \frac{|p| + 3h/2}{|p|-3h/2} \right) < -\frac{1}{2|p|} \arctanh 
\left( \frac{h}{2|p|} \right) \,, 
\ee 
which yields 
\be 
\ln \left( \sqrt{ 
\frac{
|p|+3h/2}{|p|-3h/2} } \, \right) >
\arctanh \left( \frac{h}{2|p|} \right) 
\ee 
and, since the hyperbolic 
tangent is a monotonically increasing function, 
\be 
\tanh x > \frac{h}{2|p|}  \label{urcaurca}
\ee 
where 
\be 
x \equiv \ln \left( \sqrt{ \frac{ 
|p|+3h/2}{|p|-3h/2}} \, \right) \,. 
\ee 
It is straightforward to calculate 
the right-hand side of~(\ref{urcaurca}), obtaining 
\be 
\frac{ \sqrt{ \frac{ 
|p|+3h/2}{|p|-3h/2} } -\frac{1}{ \sqrt{ \frac{
|p|+3h/2}{|p|-3h/2} }
}   }{ 
\sqrt{ \frac{ |p|+3h/2}{|p|-3h/2} } + \frac{1}{ \sqrt{ \frac{
|p|+3h/2}{|p|-3h/2} } } } = \frac{3h}{2|p|} \,. 
\ee 
Therefore, 
\be 
t_1<t_3  \quad \Leftrightarrow \quad \frac{h}{2|p|} < \frac{3h}{2|p|} \,, 
\ee 
which obviously is always satisfied, hence $-t_2<t_3$ for all values 
of $h>0$ and $p <-3h/2$.

\section{Proof that $t_1<t_3$} \label{Appendix:B} 
\renewcommand{\theequation}{B.\arabic{equation}} \setcounter{equation}{0}

The inequality $t_1<t_3$ is equivalent to \be \frac{1}{4p} \, \ln \left( 
\frac{p-3h/2}{p+3h/2} \right) < -\frac{1}{2p} \arctanh \left( \frac{h}{2p} 
\right) \ee which, since $p>0$, becomes 
\be 
\arctanh \left( \frac{h}{2p} 
\right) < \ln \left( \sqrt{ \frac{ p+3h/2}{p-3h/2} } \, \right) \,, \ee or 
\be 
\frac{h}{2p} < \tanh y \label{Bquesta} 
\ee 
where 
\be 
y \equiv \ln  \left( \sqrt{ \frac{ p+3h/2}{p-3h/2}} \, \right) \,. 
\ee 
The right-hand side of Eq.~(\ref{Bquesta}) becomes 
\be 
\frac{ \sqrt{ \frac{  p+3h/2}{p-3h/2} } -\frac{1}{ \sqrt{ \frac{ 
p+3h/2}{p-3h/2} } }   }{
\sqrt{ \frac{ p+3h/2}{p-3h/2} } + \frac{1}{ \sqrt{ \frac{ p+3h/2}{p-3h/2}
} }
}
= \frac{3h}{2p} \,, 
\ee 
giving 
\be 
t_1<t_3 \quad \Leftrightarrow \quad 
\frac{h}{2p} < \frac{3h}{2p} \,, 
\ee 
which obviously is always satisfied since both $h$ and $p$ are positive. 
Therefore, it is $t_1<t_3$ for all values of $h>0$ and $p>3h/2$.

\end{appendices}

\end{document}